\documentclass[final]{ustcstep}
\usepackage[dvips]{graphicx}
\usepackage[square]{natbib}

\def\gsim{\;\lower4pt\hbox{${\buildrel\displaystyle >\over\sim}$}\;}
\def\lsim{\;\lower4pt\hbox{${\buildrel\displaystyle <\over\sim}$}\;}
\def\grls{\;\lower4pt\hbox{${\buildrel\displaystyle >\over <}$}\;}

\newcommand\addr[2]{{\footnotesize \it $^{#1}$#2}\\}

\begin{document}

\title{Critical Height for the Destabilization of Solar Prominences: Statistical Results from STEREO Observations}

\author{Kai Liu, Yuming Wang$^*$, Chenglong Shen, and Shui Wang\\[1pt]
\addr{}{CAS Key Laboratory of Geospace Environment,
Department of Geophysics and Planetary Sciences,}
\addr{}{University of
Science \& Technology of China, Hefei, Anhui 230026, China}
\addr{*}{To whom correspondence should be addressed. E-mail: ymwang@ustc.edu.cn}
}

\maketitle
\tableofcontents

\begin{abstract}
At which height will a prominence inclined to be unstable, or where
is the most probable critical height for the prominence
destabilization? This question is statistically studied based on 362
solar limb prominences well-recognized by SLIPCAT from 2007 April to
the end of 2009. We found that there are about 71\% disrupted
prominences (DPs), among which about 42\% of them did not erupt
successfully and about 89\% of them experienced a sudden
destabilization (SD) process. After a comprehensive analysis of the
DPs, the following findings are discovered. (1) Most DPs become
unstable at the height of 0.06 -- 0.14 R$_\odot$ from the solar
surface, and there are two most probable critical heights, at which
a prominence is much likely to get unstable; the primary one is 0.13
R$_\odot$ and the secondary one is 0.19 R$_\odot$. (2) There exists
upper limit for the erupting velocity of eruptive prominences (EPs),
which decreases following a power law with increasing height and
mass; the kinetic energy of EPs accordingly has an upper limit too,
which decreases as critical height increases. (3) Stable prominences (SPs) are generally longer and
heavier than DPs, and not higher than 0.4 R$_\odot$. (4) About 62\% of EPs were associated with CMEs;
but there is no difference in apparent properties between EPs
associated with and without CMEs.
\end{abstract}

\section{Introduction}\label{sec1}
Prominences (filaments) are singularities in the corona since they
consist of cool (temperature $T\le10^4K$) and dense (electron
density $10^9$ -- $10^{11}$ cm$^{-3}$) plasma in the hot and diluted
coronal medium\citep{Spiros2002}. They are sustained and confined by
magnetic field lines above the chromospheres, and exhibit a strong
coupling between magnetic forces and thermodynamics
\citep{Wiik1997}.

When a prominence ascends with a significant velocity, it is called
eruptive prominence \citep[EP,][]{Pettit1950}. As one of the
earliest known forms of mass ejections from the Sun, EPs started
to receive attentions since the late 1800s
\citep[See Chapter 1 in book by][]{Tandberg1995}. Many authors have studied the relationship
between EPs with coronal mass ejections (CMEs) and indicate a close
association between the two phenomena
\citep[e.g.,][]{Holly2000,Gopalswamy2003,Carolus2008,Filippov2008_1}.
\citet{House1981} suggested that the inner core of CMEs is made up
of prominence material, which is believed to be the remnants of EPs.
So EPs can be treated as a tracer of CMEs \citep{Engvold2000}, and
the study of the kinematic evolution of EPs may advance our ability
in prediction of the launch of CMEs.

According to \citet{Zirin1979}, prominences are inclined to erupt
when their heights exceed 50 Mm. We know that the size and height of
a prominence increase with age \citep{Rompolt1990} and the
prominence height characterizes the surrounding magnetic field
\citep{Makarov1992}, which is crucial to the stability of
prominences. Based on the inverse-polarity model
\citep{Kuperus1974}, \citet{Filippov2000} deduced that, by 
assuming the change of magnetic field in height a power-law 
function, a quiescent prominence is stable if the power index
is less than unit, and will erupt when the power index
becomes and exceeds unit. The magnetic field in corona at low 
heights is nearly homogeneous and decreases as the field of a 
bipole ($h^{-3}$) at higher heights, thus the
equilibrium of a prominence would be unstable when it reaches a
certain height. This height is critical for the destabilization of
prominence; we call it critical height in brief.
According to \citet{Filippov2000} paper, its value can be calculated by
\begin{eqnarray}
h_c=-\frac{B}{dB/dh|_{h_c}} \label{equ1}
\end{eqnarray}
where $B$ is the magnetic field strength, $h$ is the height of 
prominence.
A follow-up study by \citet{Filippov2008_2} did show that prominences erupted near the
heights calculated by Eq.~\ref{equ1}.

In this paper, we will re-exam the historical issue statistically by
using the EUV 304{\AA} data from STEREO EUVI instrument. STEREO has
the so far most complete and uninterrupted observations at EUV
304{\AA} wavelength since the late 2006. It gives us a chance to
study the critical height for prominence destabilization in a
statistical way with a large sample. We define that a disrupted
prominence (DP) is a prominence destabilized during the period of it
being detected, and a stable prominence (SP) is the prominence else.
In Section~\ref{sec2}, how we pick and classify prominences for our
study is introduced. The statistical results are given in
Section~\ref{sec3}. Section~\ref{sec4} is designed for the
conclusions and discussions.

\section{Data}\label{sec2}
\subsection{Selection of Prominences}
Prominences above the solar limb can be clearly detected at EUV
304{\AA} wavelength. In our previous work, a system called SLIPCAT
(Solar LImb Prominence CAtcher \& Tracker) had been developed to
recognize and track solar limb prominences based on only He II
304{\AA} data observed by SECCHI/EUVI onboard of STEREO. The
technique of region-growing with thresholds and linear discriminant
analysis are two major functions applied by SLIPCAT to recognize
prominences. Lots of limb prominences as well as various parameters
of each recognized prominence are obtained, and an web-based online
catalog has been generated (\citealt{Wang2010}, hereafter Paper I).
Now SLIPCAT has a complete data set for both STEREO--A and STEREO--B
data from 2007 April to 2010 April
(http://space.ustc.edu.cn/dreams/slipcat/).

In this paper, we will use the data set of STEREO--B from 2007 April
to the end of 2009 for our study. We use STEREO--B rather than
STEREO--A because STEREO--B has a larger field of view (FOV) than
STEREO--A. We do not involve the data after January 2010, because
the solar activity level obviously increased, and the set of
calculation parameters used by SLIPCAT is different from that for
the STEREO data before 2010 (refer to our website for more details).

\begin{figure}[tbh]
  \centering
  \includegraphics[width=\hsize]{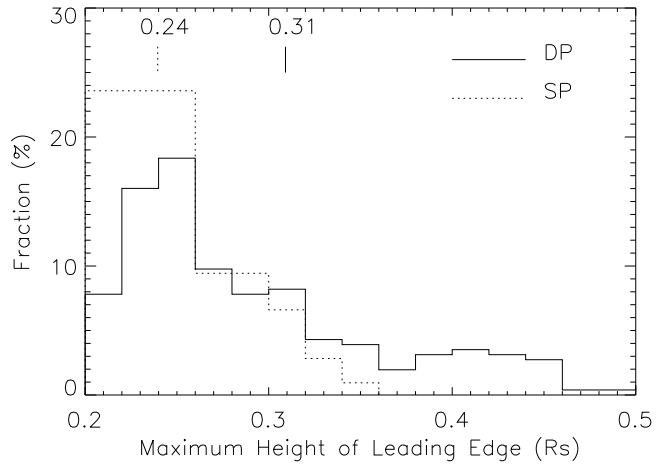}
  \caption{\label{fig01}Distribution of the maximum height
of leading edges of DPs (solid line) \& SPs (dotted line). The short
vertical lines with digital numbers mark the average values of histograms.}
\end{figure}

SLIPCAT extracted 10072 `well-tracked' (see Paper I for the
definition of the term) prominences based on SECCHI/EUVI 304{\AA}
data from STEREO-B during the period of interesting. Further, we
narrow our sample by selecting prominences whose maximum height of
leading edge from solar surface is greater than $h=0.2$ R$_\odot$ (or
$140$ Mm). We believe that most prominences below 0.2 R$_\odot$ are
not eruptive. A reason is that \citet{Munro1979} found that all
eruptive prominences observed beyond 0.2 R$_\odot$ were associated
with CMEs, which is consistent with \citet{Holly2000}'s study, in
which it is found that all 18 EPs with a maximum height greater than
0.2 R$_\odot$ had an associated CME and only one failed to reach 0.2
R$_\odot$. Moreover, a stronger reason is given in
Figure~\ref{fig01}, which shows the distributions of the maximum
height of the leading edges of SPs and DPs (the identification
process of SPs and DPs is described in Sec.~\ref{sec2.2}). For SPs,
there are more event number at lower height, while for DPs, there is
a peak appearing at the middle of the distribution. It indicates
that the selection of 0.2 R$_\odot$ can cover most DPs.

Besides, it should be noted that the prominence height obtained
by SLIPCAT is projected one, and therefore is underestimated. Since
prominence is an extended structure, the underestimation will be 
even large for some prominences locating at a significant distance
away from limb. For a prominence with its top locating in 20 degrees
away from limb, it could be derived that its height be 
underestimated by $0.06(h+1)$ R$_\odot$, which is less than 0.1 R$_\odot$
in our sample. Hence, we believe that most 
detected prominences should not be too far away limb, and therefore 
the projection effect will not significantly affect our statistical
results.

\subsection{Classifications of Prominences}\label{sec2.2}
\begin{figure*}[tbh]
  \centering
  \includegraphics[width=\hsize]{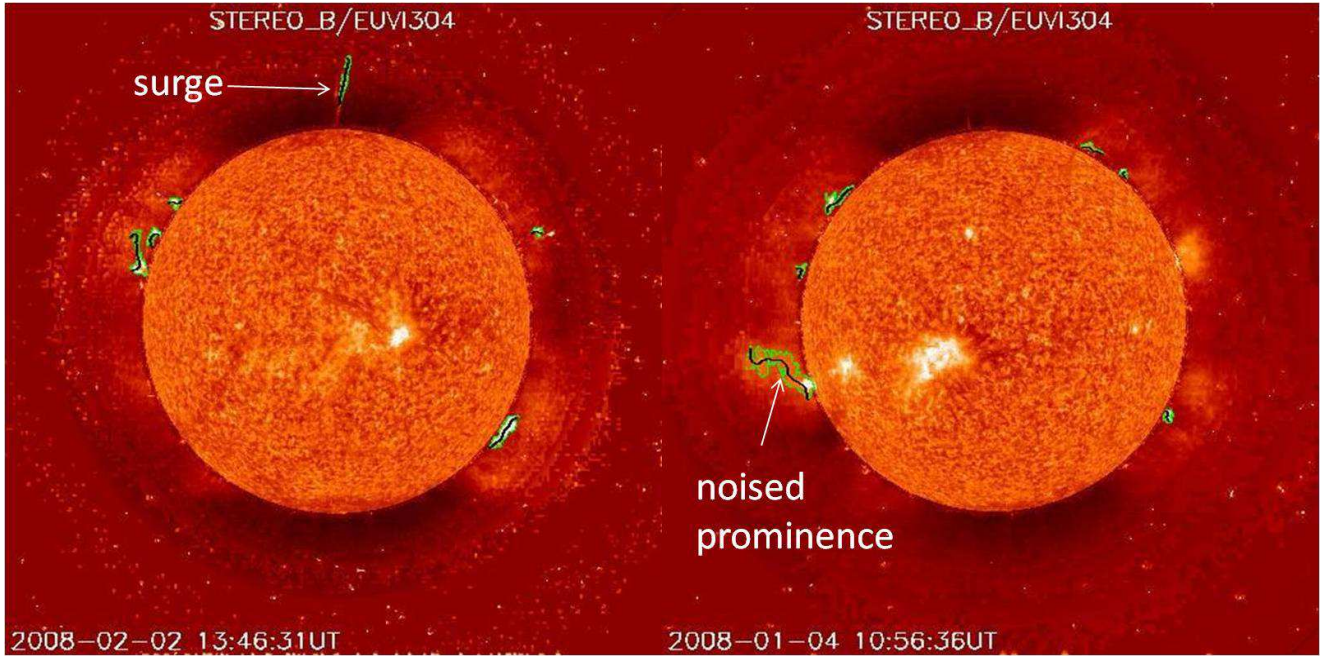}
  \caption{\label{fig02}Example showing a surge (left)
and a noised prominence (right).}
\end{figure*}

Although almost all prominences can be recognized by SLIPCAT,
however, it should be admitted that not all the recognized
prominences are real prominences. Some of them are surges, and some
are not well-recognized but contaminated by noise (Fig.\ref{fig02}).
Surges have a much different appearance and behavior from a typical
prominence. They are relatively narrow and faint, and usually erupts
quickly and radially, which means a short lifetime. By manually
checking EUV 304{\AA} movies, we remove all surges and noised
prominences from our sample. Meanwhile, all the real prominences are
classified into different types according to their dynamic processes
during their being detected, which have been listed in
Table~\ref{t1}.

\begin{figure*}[p]
  \centering
  \includegraphics[width=\hsize]{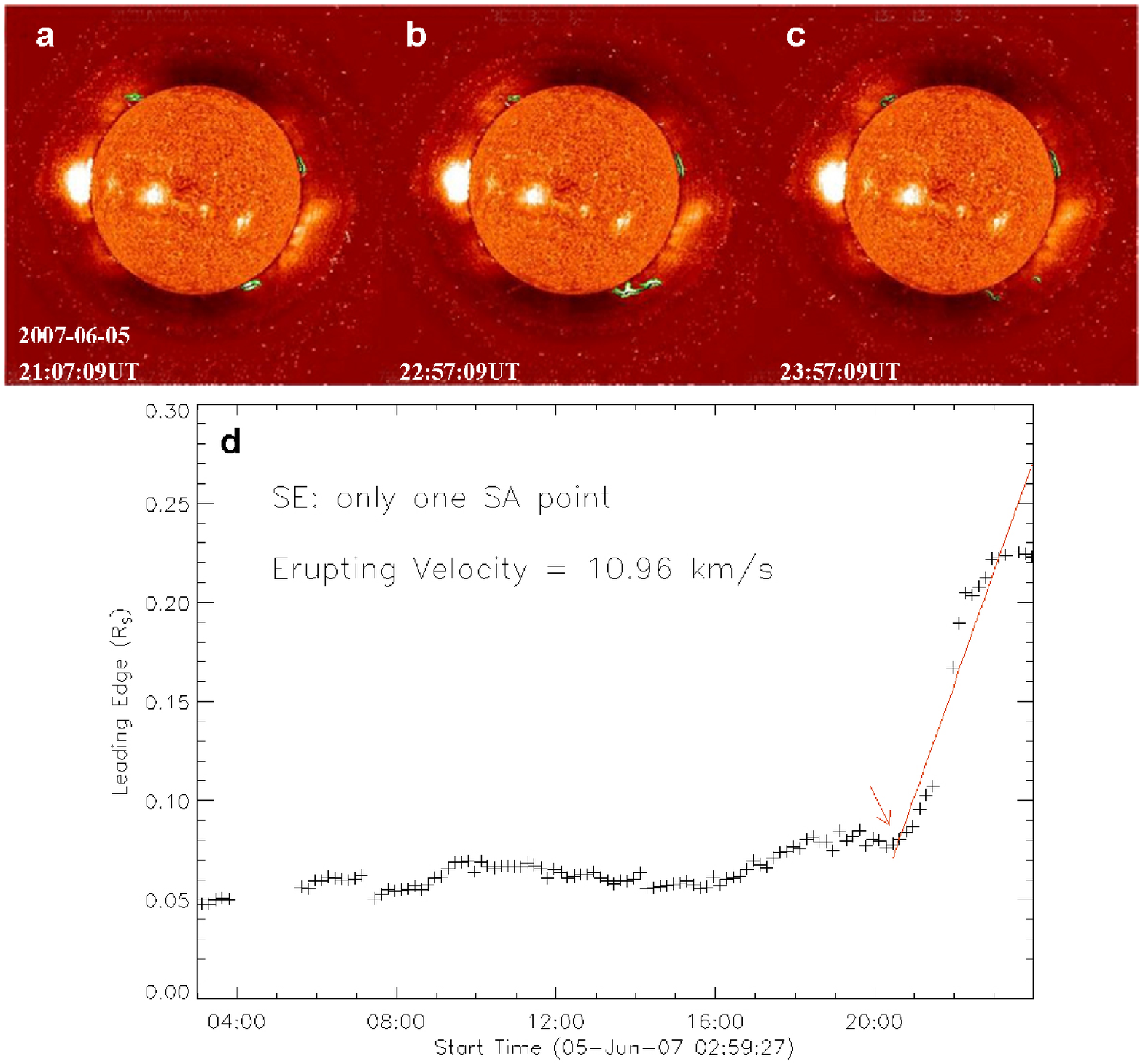}
  \caption{\label{fig03}An example of SE.}
\end{figure*}

\begin{figure*}[p]
  \centering
  \includegraphics[width=\hsize]{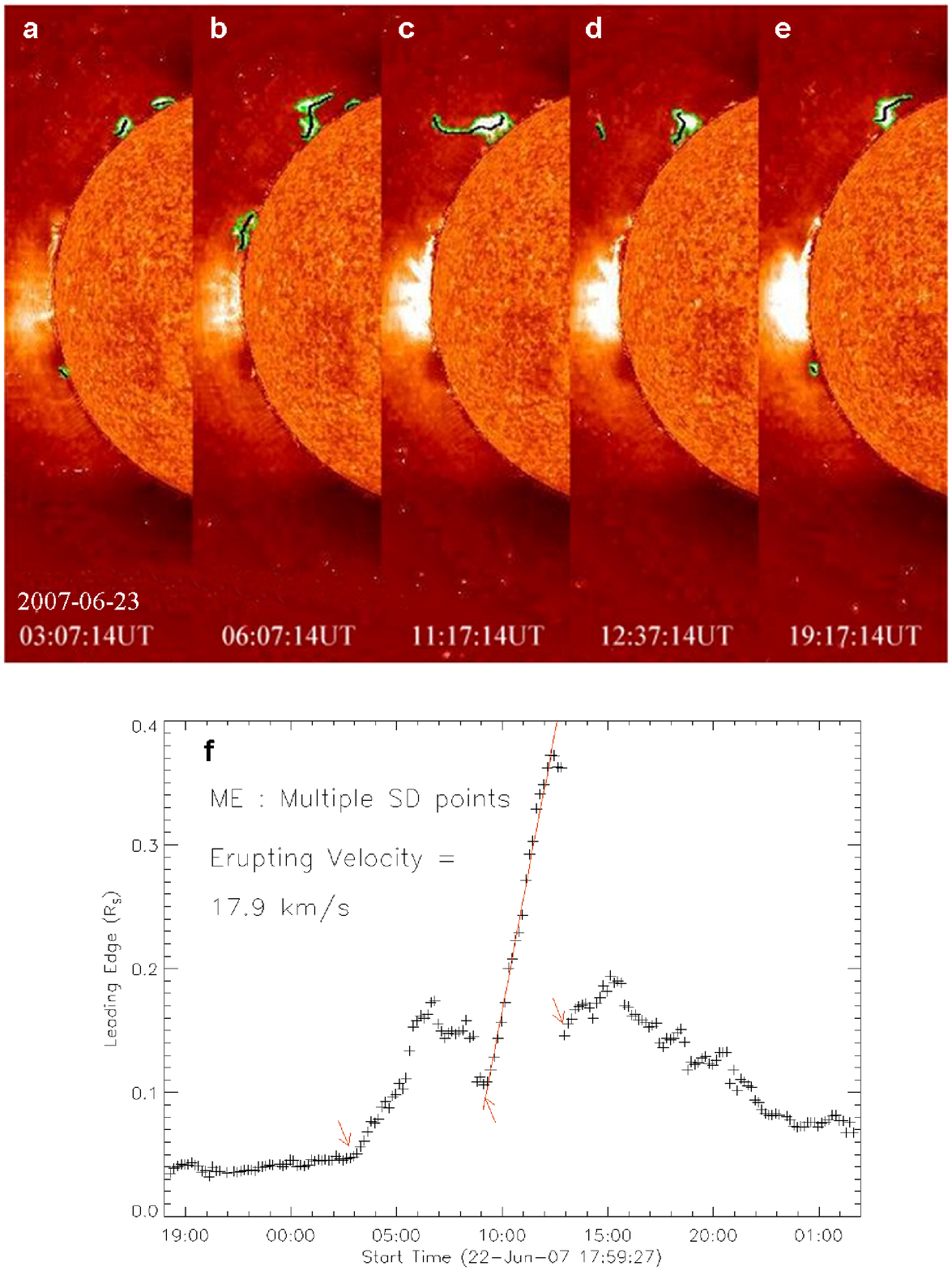}
  \caption{\label{fig04}An example of ME.}
\end{figure*}

\begin{figure*}[p]
  \centering
  \includegraphics[width=\hsize]{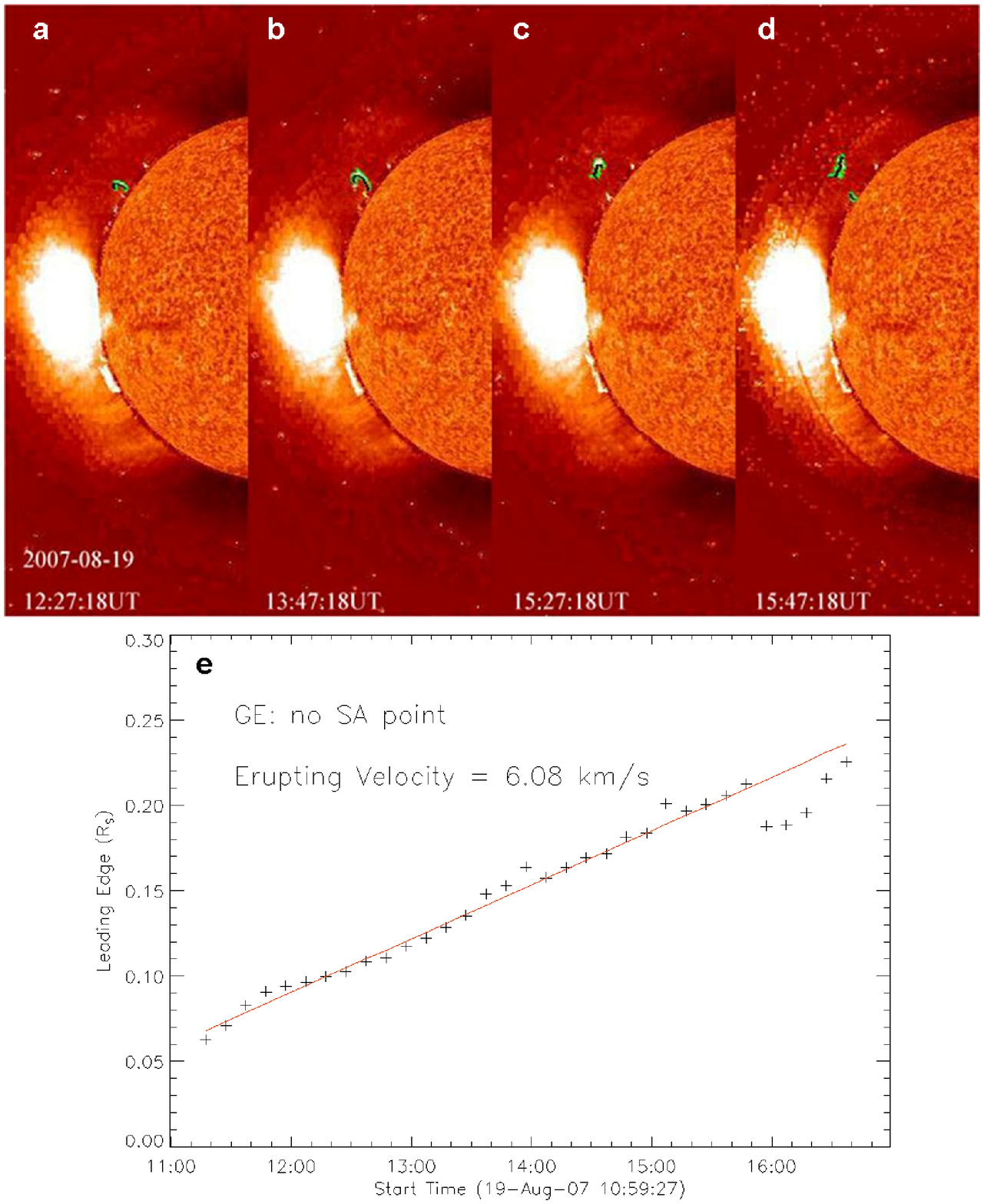}
  \caption{\label{fig05}An example of GE.}
\end{figure*}

First we classified prominences into two main types as SPs and DPs,
which have been defined in the end of Sec.1. Second, DPs are
classified as EPs and failure eruptive prominences (FPs). We use the
definition from \citet{Holly2000}'s study to define an EP as one in
which all or some of the prominence material appears to escape the
solar gravitational field\footnote{Such apparently escaping may be
real, but also possibly be faked due to some thermodynamic processes
(See Holly et al. 2000 and Paper I). In this paper, we do not
distinguish an EP does erupt or is just heated.}. If the ascending
prominence material apparently falls back down in the FOV of
SECCHI/EUVI ($\leq1.7$ R$_\odot$ for STEREO--B), we called these
prominences FPs.

As an EP evolves, its leading edge may be gradual rising or
experience one or multiple sudden destabilization (SD) phases. The
latter means that there could be one or several points, at which the
prominence get destabilization. Thus, we further divide EPs into
three types. The first type is EPs with only one SD phase, which we
called single-phase eruption (SE). Figure~\ref{fig03}a--\ref{fig03}c
are three images processed by SLIPCAT showing a SE right before the
eruption, during the eruption, and before it faded away around the
position angle of about 218$^\circ$ (near the South Pole).
Figure~\ref{fig03}d gives the profile of the entire evolution
process of its leading edge since 2007-06-05 03:07 UT. It is clear
that there is one obvious SD point (denoted by the red arrow),
before which the prominence slowly rose with a weak oscillation, and
after which it was obviously suddenly accelerated to get a
significant speed. The SD point just indicates the critical height
of the destabilization of the prominence, and implies a catastrophic
process. The erupting speed is estimated to be about 11 km/s by
using a linear fitting.

Second type is more complicated. It contains EPs with multiple SD
phases, which we called multiple-phase eruption (ME).
Figure~\ref{fig04} shows an EP with position angle around 37$^\circ$
on 2007 June 23 for example. There are three SD points (marked in
the Fig.\ref{fig04}f as red arrows) corresponding to three SD phases
in this EP's evolution process. The first and third SD phase both
indicate a failed eruption, which can be seen from
Figure~\ref{fig04}a--\ref{fig04}b and
Figure~\ref{fig04}d--\ref{fig04}e. The second SD phase indicates a
successful eruption, which is can be seen from
Figure~\ref{fig04}b--\ref{fig04}d, and that is the reason we
classified this prominence as an EP. We choose the height of the
first SD point of ME as its critical height, because that is the
point where the prominence loses its stability for the first time.
We obtained the eruption speed of this EP by doing a linear fitting
to the profile of the successful eruption phase, which is from the
second SD point to the point where the escaping material is fading
away, and the value is about 18 km/s.

When an EP's leading edge is rising gradually, we call it as gradual
eruption (GE), which is the third type of our classification. We
also give an example of GE in Figure~\ref{fig05}.
Figure~\ref{fig05}a--\ref{fig05}d demonstrate its appearance through
fading away around position angle of 59$^\circ$. Being different
from the above two types, a GE does not have a SD point at all, as
seeing from the evolution profile in Figure~\ref{fig05}e. Thus, we
ca not get the critical height for the destabilization of this type
of EP. The ascending speed of the GE event is about 6 km/s by using
a linear fit.

FPs also have one or multiple SD phases just as EPs, although the
erupting materials of FPs fail down eventually according to our
data, they also have critical heights for their SD phases. As same
as EPs, we take the height of the first SD of a FP as its critical
height. Thus, in our classification, SEs, MEs and FPs can be found a
critical height where they destabilized.

In summary, there are a total of 362 well-recognized prominences
with the maximum height of leading edge above 0.2 R$_\odot$ detected
by STEREO--B during 2007 April -- 2009 December (Table~\ref{t1}). In
these events, there are 106 (occupying 29\%) SPs during the period
of detection and 256 (occupying 71\%) DPs. The latter consists of
107 (42\%) FPs and 149 (58\%) EPs, and EPs further consist of 70
(47\%) SEs, 50 (34\%) MEs and 29 (19\%) GEs. Besides, we list all
the acronyms in the appendix for one's reference.

\begin{table}[tbh]
\centering
\caption{\label{t1}Classification of prominences}
\begin{tabular}{c|ccc|c|c}
  \hline
   \multirow{3}{*}{SPs} & \multicolumn{4}{c|}{DPs} & \multirow{3}{*}{Total} \\
   \cline{2-5}
   & \multicolumn{3}{c|}{EPs} & \multirow{2}{*}{FPs} & \\
   \cline{2-4}
   & SEs &  MEs & GEs & &  \\
   \hline
   106  & 70 & 50 & 29 & 107 & 362 \\
   29\% & 19\%  & 14\%  & 8\%   & 30\%  & 100\% \\\hline
\end{tabular}
\end{table}

\section{Results}\label{sec3}
\subsection{Distribution of Critical Heights}

\begin{figure}[tbh]
  \centering
  \includegraphics[width=\hsize]{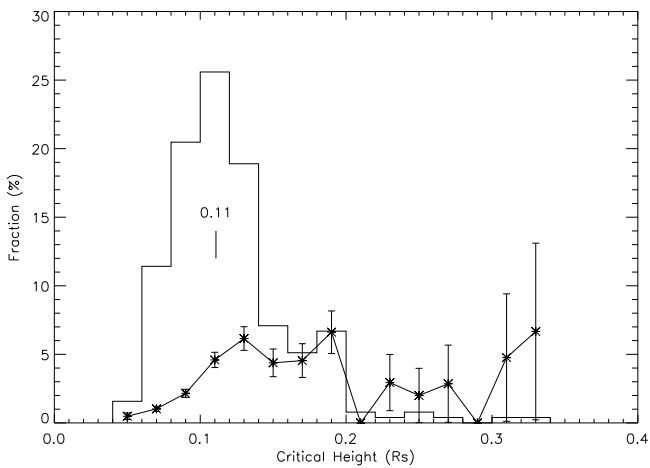}
  \caption{\label{fig06}Distribution of critical heights (histogram) and possibilities of prominences becoming unstable at various heights (asterisks connected with lines). See main text for more details.}
\end{figure}

The distribution of the critical heights of EPs (except GEs) and
FPs, which contains 227 data points, is shown in Figure~\ref{fig06}.
From the histogram, it is found that about 76\% of the critical
heights fall in the range of 0.06 -- 0.14 R$_\odot$ with the mean
value at around 0.11 R$_\odot$. Further, the asterisks connected
with lines show the ratio of the number of SEs, MEs and FPs to the
number of all recognized prominences by SLIPCAT (for those
prominences without a critical height, the maximum height of leading
edge is used). The value of the asterisk actually indicates the
possibility of being unstable when a prominence reaches a certain
height. The uncertainty is marked by the error bars, which
calculated by the formula $\sigma=[p(1-p)/N]^{1/2}$, where $p$ is
the possibility, $N$ is the total number of events in the bin. It
should be mentioned that, the surges and noised prominences with the
maximum height of leading edge larger than 0.20 R$_\odot$ are
removed, but not for those smaller than 0.20 R$_\odot$. However,
after a quick exam of EUVI movies, we found that most surges and
noised prominences have maximum leading edge greater than 0.20
R$_\odot$ and therefore the inclusion of surges and noised
prominences with the maximum leading edge lower than 0.20 R$_\odot$
will not significantly affect the values of the possibilities. The
possibility distribution above the height of 0.22 R$_\odot$ is not
reliable, because the event number is small and the uncertainty is
significantly large. The small fraction of the events above the
height of 0.22 R$_\odot$ implies that the most probable critical
height will not be there. The distribution below the height of 0.22
R$_\odot$shows a double-peak feature. The two peaks appear at the
height of 0.13 and 0.19 R$_\odot$, respectively. Since the first one
falls in the range of 0.06 -- 0.14 R$_\odot$, in which there are
76\% of critical heights, we conclude that 0.13 R$_\odot$ is the
primary most probable critical height and 0.19 R$_\odot$ is the
secondary one.

\begin{figure}[tbh]
  \centering
  \includegraphics[width=\hsize]{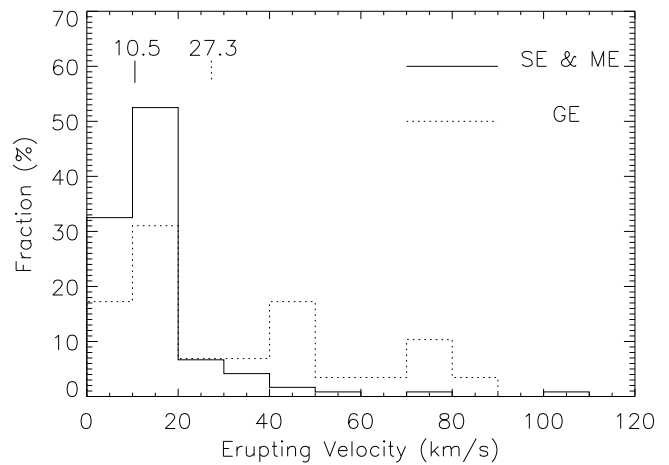}
  \caption{\label{fig07}Distribution of EP's erupting velocity. The vertical lines with digital numbers indicate the average values.}
\end{figure}

GEs do not have a SD point though they erupted too. Whether or not
their erupting velocities are systematically different from others?
This issue is inspected by comparing the distributions of the
erupting velocities of GEs and the other two types of EPs (Fig.\ref{fig07}).
The average velocity of all EPs is about 14 km s$^{-1}$.
It is found that GEs tend to have a larger erupting velocity. The
result is apparently contrary to the usual physical picture that an
impulsive eruption (i.e., there is a SD point) should be faster than
a gradual eruption. An explanation we can have right now is
that those fast GEs probably already have passed through their
critical heights before they are recognized by SLIPCAT. It is possible
if most of the GEs located at a significant distance away from limb.
They will not be noticed until they have risen or erupted to exceed 
the limb.


\subsection{Erupting Velocity of EPs}\label{sec3.2}
\begin{figure*}[p]
  \centering
  \includegraphics[width=\hsize]{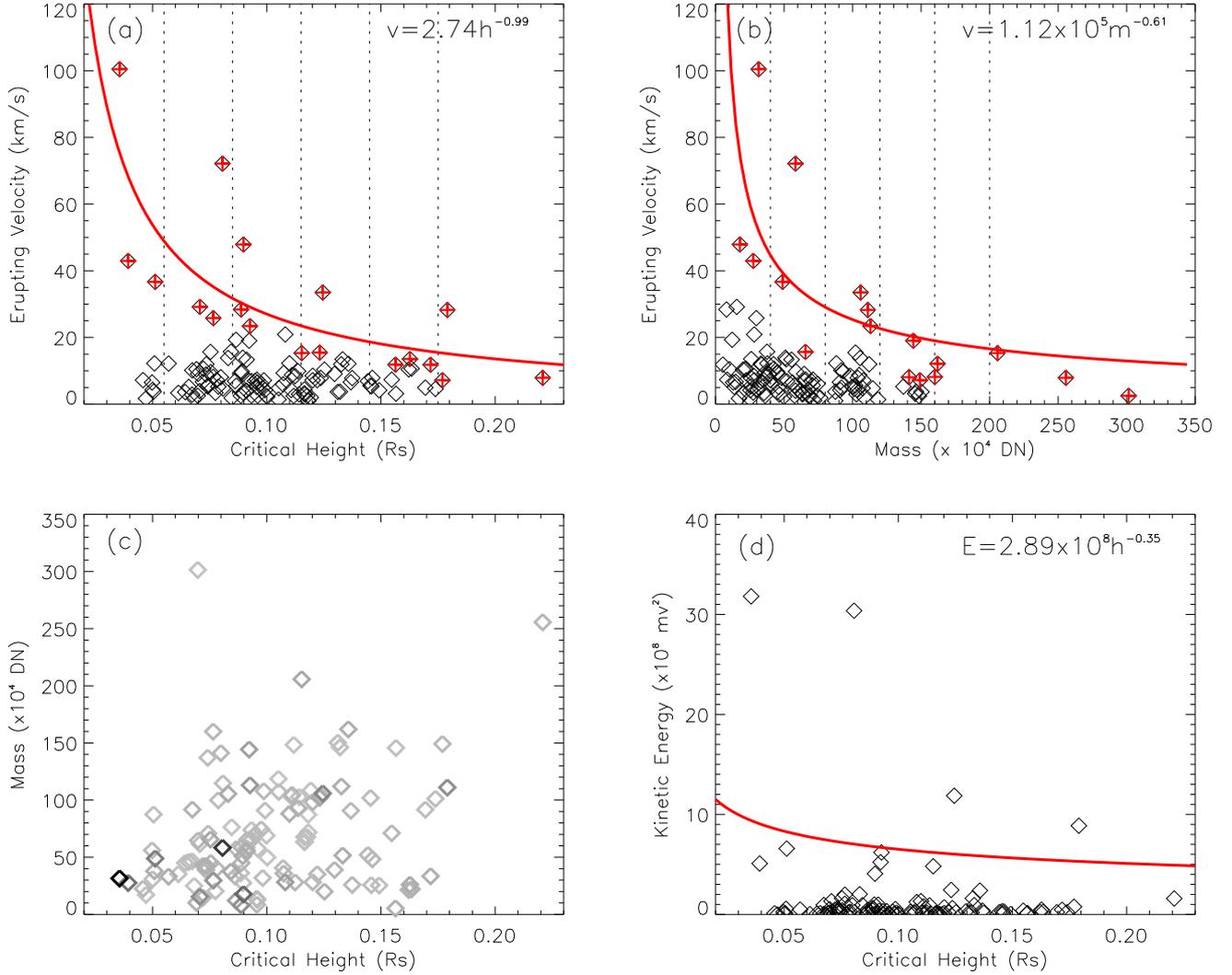}
  \caption{\label{fig08}
Correlation between (a) erupting velocity and critical height,
(b) erupting velocity and mass (represented by total brightness),
(c) mass and critical height, and (d) kinetic energy and critical
height. In panel (a) and (b), the dashed vertical lines separate
data points into 6 bins, plus symbols mark the data points 
having top three erupting velocity in every bin, and the solid 
line is the fitting curve to them, which is given by the formula
at the upper right corner. In panel (c) the darker points have 
larger erupting velocities. In panel (d) the solid line, given by
the formula at the upper right corner, is derived from the
fitting functions in panel (a) and (b).}
\end{figure*}

The comparison between the erupting velocity of GEs and the
other two types of EPs has been presented in the last section.
Here we will further investigate the erupting velocity of SEs
and MEs and its correlations with critical height and mass.
In this study, mass is approximated as the total brightness 
recorded by STEREO/EUVI at 304\AA\ wavelength, and in units of 
digital number (DN). Scattering plots of the erupting velocity
versus critical height and the erupting velocity versus mass are
shown in Figure~\ref{fig08}a and \ref{fig08}b. It is obvious that
the maximum erupting velocity decreases with increasing critical
height and mass. This result suggests that, at a certain height or
for a certain mass, the kinetic energy an eruptive prominence
could reach or the free magnetic energy accumulated in the
prominence-related magnetic system has an upper limit.

In order to obtain the dependences of the upper limit on the
height and mass, we divided the data sample into six bins as
indicated by the equally-separated vertical dashed lines. The
data points with erupting velocity at the top three position in each
bin (marked as plus symbols) 
are fitted by a function $v=c_0x^{c_1}$. For
Figure~\ref{fig08}a, $x$ is $h$ the critical
height calculated from solar surface; for Figure~\ref{fig08}b,
$x$ is just the mass represented
by the total brightness. The fitting results give the following
two equations
\begin{eqnarray}
v_{max}&=&2.74h^{-0.99} \mathrm{\ (km\ s^{-1})}\\
v_{max}&=&1.12\times10^5m^{-0.61} \mathrm{\ (km\ s^{-1})}
\end{eqnarray}
in which $h$ is in units of R$_\odot$ and $m$ in units of DN.
Figure~\ref{fig08}c shows the mass versus the critical height. The 
darker symbol means a larger velocity. Consistent with the above
results, the fast erupting prominences locate at the lower-left 
corner. No obvious correlation between the two parameters is revealed, which
indicates that the correlations of the maximum erupting velocity
with the critical height and mass are almost independent.

Further, we can derive the upper limit of kinetic energy of 
eruptive prominences as a function of critical height from the 
above fitting results as the follow equation
\begin{eqnarray}
E_{max}\propto mv_{max}^2=2.89\times10^8h^{-0.35} \mathrm{\ (DN\ km^2\ s^{-2})}\label{eqeh}
\end{eqnarray}
The red line in Figure \ref{fig08}d presents the equation.
Comparing it with the data points in that plot, we can 
find a weak consistency between them with only four data points 
exceeding the upper limit.

\begin{figure*}[tbh]
  \centering
  \includegraphics[width=\hsize]{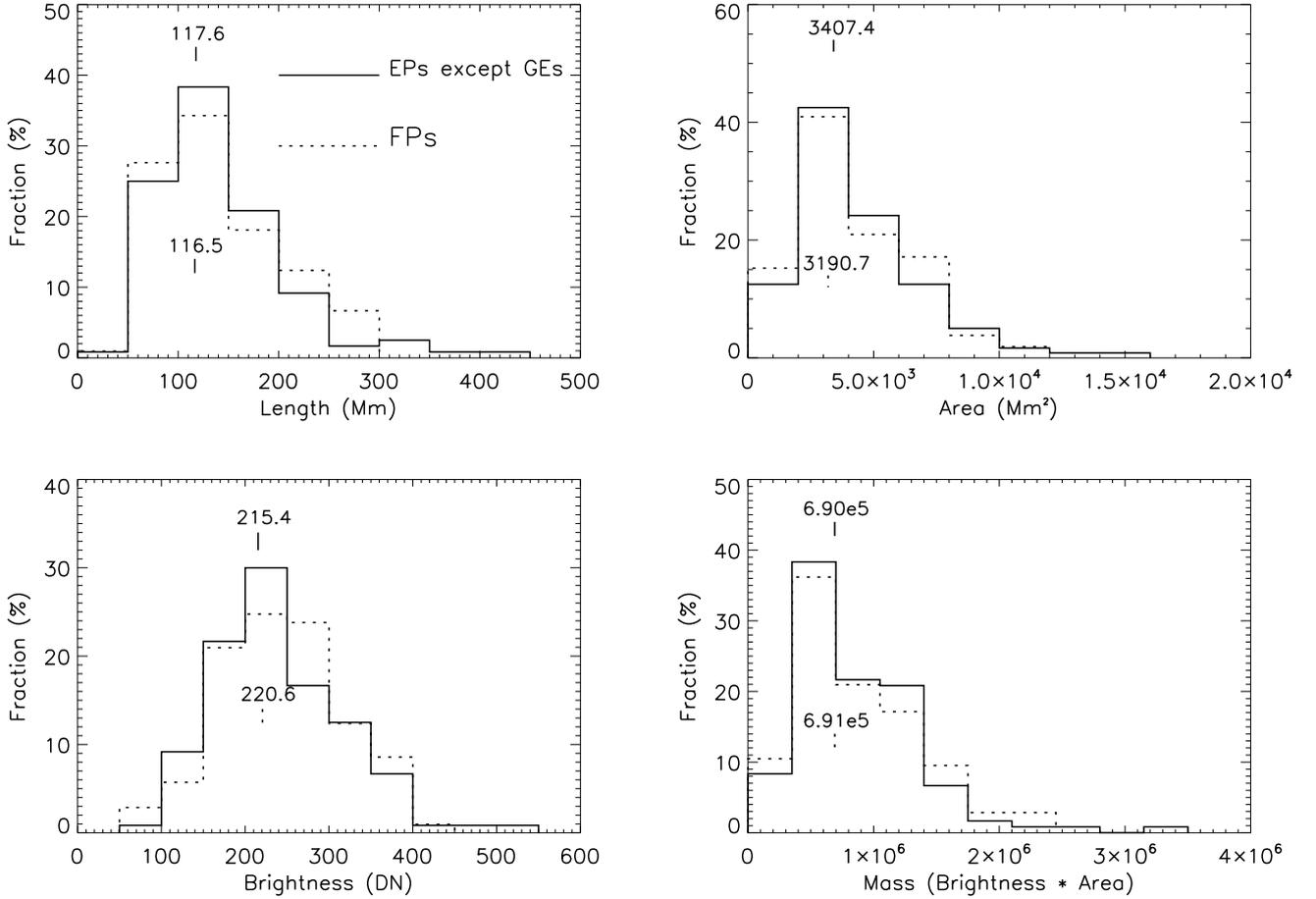}
  \caption{\label{fig09}Distributions of the apparent properties of
EPs except GEs (solid lines) and FPs (dotted lines) at SD point.}
\end{figure*}

\subsection{EPs vs. FPs \& DPs vs. SPs}
Although both EPs (except GEs) and FPs have a SD point, is
there crucial difference between the properties of these two
types of DPs when they go through SD point? With this
question, we compared their length, area, brightness and
mass at SD point, and plotted them as histograms in
Figure~\ref{fig09}. It is shown that there is no obvious
difference between the two types, which indicates that one
can not use these properties of a DP at SD point to decide
whether or not it would have a successful eruption.

\begin{figure*}[tbh]
  \centering
  \includegraphics[width=\hsize]{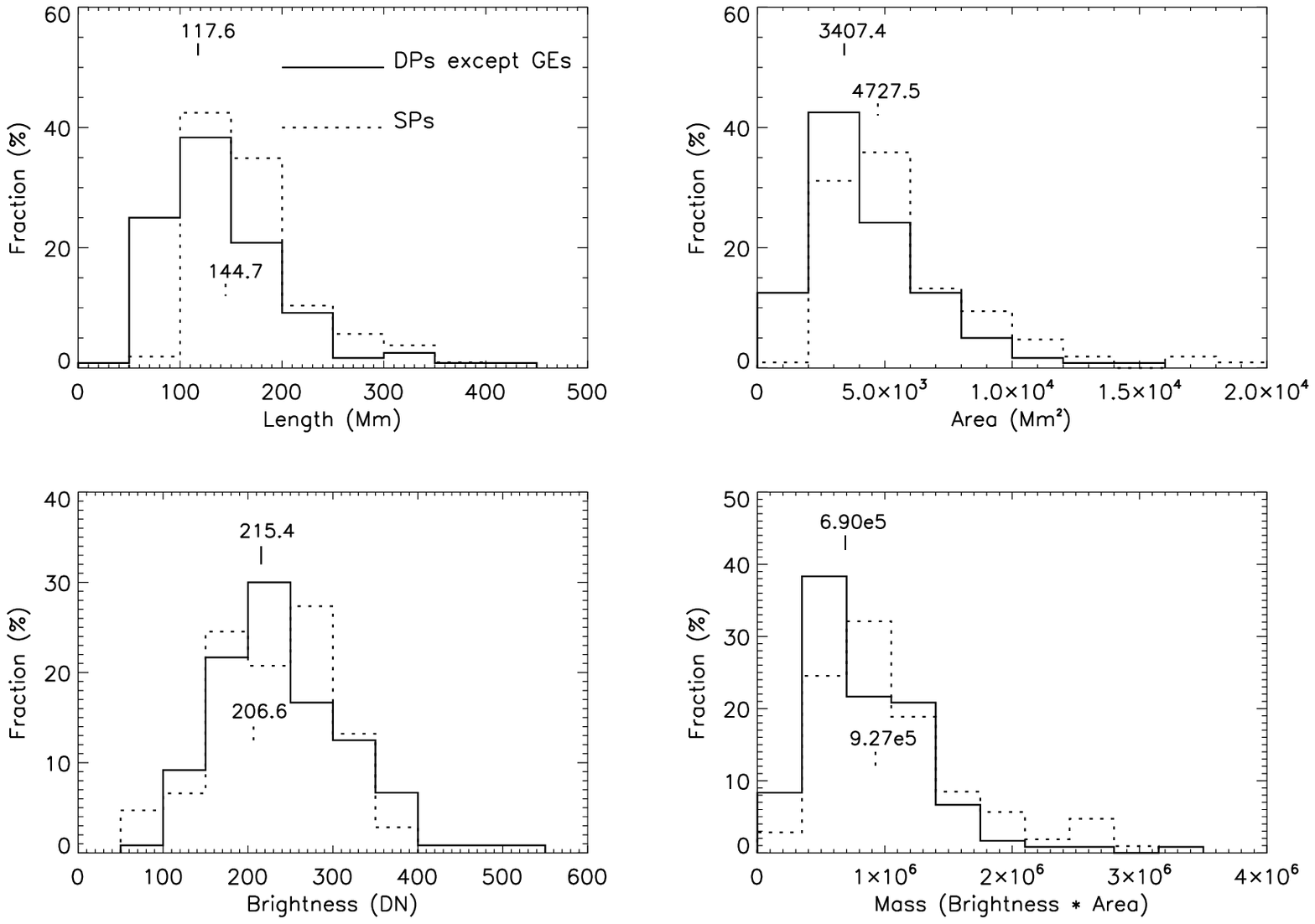}
  \caption{\label{fig10}Distributions of the apparent
properties of DPs except GEs at SD point (solid lines) and the average values
of the apparent properties of SPs (dotted lines).}
\end{figure*}

SPs do not have much change in their properties during their
appearances, so we can use the mean values of the length, area,
brightness and mass to present them. We compared these mean values
with DPs' properties at SD point (Fig.\ref{fig10}) to check if there
is any difference. The histograms in Figure~\ref{fig10} show that
SPs generally have a larger length, area and mass than DPs by a
factor of about 1.3. It is reasonable that a large and heavy
prominence tends to be stable. It is also consistent with the
results obtained in the last section that a prominence with larger
mass would have a smaller upper limit of erupting velocity.

\begin{figure*}[tbh]
  \centering
  \includegraphics[width=\hsize]{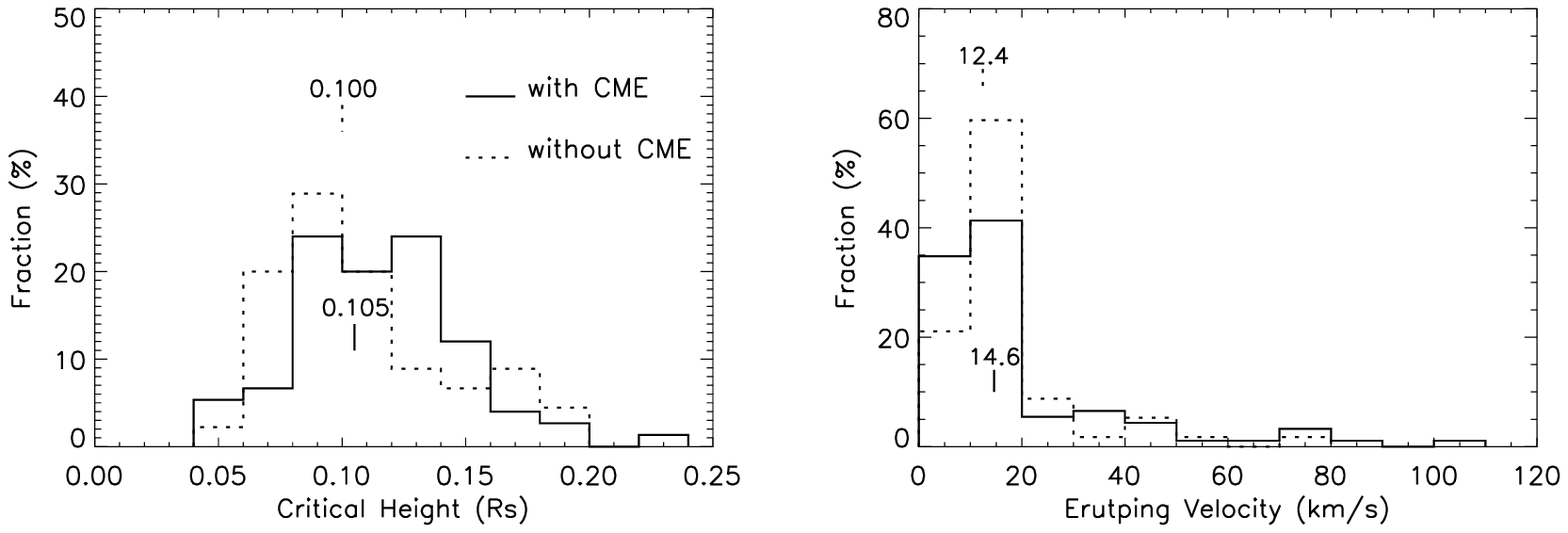}
  \caption{\label{fig11}Distributions of critical
height (left) and erupting velocity (right) of EPs with
(solid lines) and without (dotted lines) CMEs.}
\end{figure*}

\subsection{CME Association of EPs}
We checked the association of EPs with CMEs by using the same
method as \citet{Holly2000}. First, we determined the position
angle and eruption time of an EP of interest. Then we browsed
the movies of both coronal observations from COR1 and COR2 on
board STEREO--B within 2 hours of the eruption time. If there
was a CME with central position angle within 30$^\circ$ of the
position angle of the EP in either coronagraph, this EP is
considered to be associated with a CME. The results are summarized
in Table~\ref{t2}.

\begin{table}[tbh]
\begin{center}
\caption{\label{t2}Association of EPs with CMEs}
\begin{tabular}{ccc}
 \hline
    & CME-assoc./Total    & Fraction \\
\hline
 SEs    & 40/70                & 57\%     \\
 MEs    & 35/50                & 70\%     \\
 GEs    & 17/29                & 59\%     \\
\hline
 Total  & 92/149                & 62\%     \\
\hline
\end{tabular}
\end{center}
\end{table}

We find that 62\% of EPs are associated with CMEs, and MEs have a
relatively higher association rate than the other two types of EPs.
The association rate we got is different from previous works, e.g.,
94\% in \citet{Holly2000} and 36\% in \citet{Yang2002}. The reason
for these differences is mostly due to the data selection.
\citet{Holly2000} used ground-based observations, and picked the
prominences that had violent change as their objects. We believe
that such a strict selection caused the high association rate.
However, \citet{Yang2002} got a much smaller association rate than
others. It is because they treated prominences with transverse
motion as EPs too, which are not considered in our sample.
\citet{Gopalswamy2003} got an association rate of about 72\%. In
their work, the data from Nobeyama Radioheliograph during 1996
January -- 2001 December was used to identify prominences. Their
study covered the period from 1996 January -- 2001 December, i.e.,
from solar minimum to maximum, whereas ours is near solar minimum.
Thus their association rate is reasonably larger than ours because
CMEs are less frequent in solar minimum than solar maximum.

We plotted distributions of EPs' critical heights
and erupting velocities with (solid line) and without
(dotted line) CMEs in Figure~\ref{fig11}. They do not
show much difference, which suggests that we are probably
not able to forecast CMEs' occurrence based only on these properties of EPs.

Figure \ref{fig11} shows that the average speed of CME-associated 
EPs is about 14.6 km s$^{-1}$, that is much lower than CME 
speed, which is typically hundreds kilometers per second in the outer corona.
This divergence implies that prominence materials can be accelerated
to considerable high speeds over one or more solar radii. The speeds
measured in EUVI FOV are just initial speeds of prominence eruptions.
Also the erupting speeds are much lower than the gravitational escape
velocity of $\sim600$ km s$^{-1}$ at the base of corona, and no obvious
deceleration could be found in the velocity profiles. This suggests a 
continuous driving from released magnetic free energy.

\section{Conclusions and Discussions}\label{sec4}
All the prominences with leading edge greater than 0.2 R$_\odot$
recognized by SLIPCAT during 2007 April -- 2009 December are
investigated. By manually examining the movies, these prominences are
classified into various types: SP, DP, EP, FP, SE, ME and GE. The
following statistical results are obtained.
\begin{enumerate}
\item There are about 71\% DPs, about 42\% of them did not erupt
successfully, and about 89\% of them experienced a SD process.

\item Most DPs become unstable at the height of 0.06 -- 0.14
R$_\odot$. There are two most probable critical heights, at which a
prominence is much likely to get unstable; the primary one is 0.13
R$_\odot$ and the secondary one is 0.19 R$_\odot$.

\item There exists upper limit for the erupting velocity of EPs,
which decreases with increasing critical height and mass. Inferentially, 
the upper limit of the kinetic energy of EPs also decreases as 
critical height increases.

\item There is no difference in apparent properties (length, area,
brightness and mass) between EPs and FPs. However, SPs are generally
longer and heavier than DPs by a factor of about 1.3, and no SPs are
higher than 0.4 R$_\odot$.

\item About 62\% of EPs were associated with CMEs. But there is no
difference in apparent properties between EPs associated with and
without CMEs.
\end{enumerate}

According to Eq.~\ref{equ1} mentioned in Sec.\ref{sec1}, we may
deduce that prominences embedded in a region with more rapid
decrease in magnetic filed strength along height should have lower
critical height. Thus, the primary and secondary most probable
critical heights may correspond to two different types of source
regions. Prominences can form in an AR or quiet Sun region. It may
be the reason why there are the two most probable critical heights.
To verify our conjecture, the source regions of DPs and the coronal
magnetic field surrounding them must be investigated, which will be
pursued in a separate paper.

Corona is dominated by magnetic field, and therefore the kinetic 
energy of prominences is converted from magnetic free energy. 
The anti-correlation of the maximum erupting velocity of EPs with
the height suggests that the maximum free energy could be
accumulated in a prominence-related magnetic field system is
larger at low altitude than at high altitude. That means the magnetic
free energy goes down as the prominence rises up or the surrounding
magnetic field structure expands. Approximately, the maximum free 
energy when the system stays at the critical height can be related to the maximum kinetic 
energy of prominences, and described as 
\begin{eqnarray}
E_{max}=\int\frac{B_f^2}{2\mu}dx^3\propto \overline{B}_f^2h^3
\end{eqnarray}
where $\overline{B}_f$ is the average magnetic field strength corresponding to the free
energy, or approximately nonpotential component of magnetic field. 
In this equation, we implicitly assume that the size 
of the volume occupied by the magnetic field is proportional to
the cubic value of critical height.
Combining with the empirical formula Eq.\ref{eqeh},
one can easily derive
\begin{eqnarray}
\overline{B}_f\propto h^{-1.7}\label{eqbh}
\end{eqnarray}
This scaling law implies that the average strength of the nonpotential
component in a magnetic field structure weakens as the structure 
rises and expands. However, this result needs to be further justified,
because Eq.\ref{eqeh} is resulted from hundreds of events and Eq.\ref{eqbh} is 
therefore established from a statistical point of view.

The manual exam of the movies gives us an impression that all the
destabilization process except GEs happened within two data points.
One may notice that the cadence our data is 10 minutes, which means
that the catastrophic process of the prominence destabilization is
shorter than 10 minutes. A question raised naturally is how quickly
such a catastrophic process progresses. To answer this question and
study why and how a prominence lose equilibrium, observations with
higher spatial and temporal resolutions must be used. Ground-based
H$\alpha$ observations may be suitable for such studies, and
besides, SDO provides the so far best space-borne observations that
may also be able to support such studies.

Besides, the dynamic evolution of prominences can be related to 
the large scale structure in corona. The coronal cavities were studied by, e.g., \citet{Fuller_etal_2008} 
and \citet{Fuller_Gibson_2009}, which are generally believed to be
flux ropes supporting quiescent prominences.
It was found that there were no cavities taller 
than 0.6 R$_\odot$. In our study, all SPs were less than 0.4 R$_\odot$ 
(Fig.\ref{fig01}).
The cavity flux rope is kept in equilibrium by two principle forces
acting against the natural tendency for the rope to expand outward,
namely, the anchored part of the field surrounding the rope, and the
weight of prominence as well as the coronal helmet,
\citep[e.g., see the static model by][]{Low_Hundhausen_1995}. 
Hence, the maximum
height for quiescent prominences is physically related to the maximum
height of coronal cavities.

\acknowledgments{We acknowledge the use of the data from
STEREO/SECCHI, and we are also grateful to the developers of
SLIPCAT. We thank Dr. BC Low for his reading and valuable comments,
and the anonymous referee for his/her constructive comments. This 
research is supported by grants from 973 key
project 2011CB811403, NSFC 41131065, 40904046, 40874075 and 41121003, CAS 
100-talent program, KZCX2-YW-QN511 and startup fund, FANEDD 200530,
and the fundamental research funds for the central universities.}

\appendix
\section*{Appendix: Acronyms}
\noindent
AR -- Active Region\\
CME -- Coronal Mass Ejection\\
DN -- Digital Number\\
DP -- Disrupted Prominence\\
EP -- Eruptive Prominence\\
EUVI -- Extreme UltraViolet Imager\\
FOV -- Field of View\\
FP -- Failed Erupting Prominence \\
GE -- Gradual Eruption \\
ME -- Multiple Eruption \\
SD -- Sudden Destabilization \\
SE -- Single Eruption \\
SLIPCAT -- Solar LImb Prominence CAtcher \& Tracker \\
SP -- Stable Prominence \\
STEREO -- Solar TErrestrial RElations Observatory

\bibliographystyle{plainnat}
\bibliography{paper}

\end{document}